\documentstyle [12pt] {report}
\begin{document}
\setlength{\baselineskip}{24pt}

\title{\large \bf  STRONG ENERGY CONDITION IN $R + R^2$ GRAVITY}

\author{\bf J. H. Kung \\
(jkung@abacus.bates.edu)\\
Department of Physics\\
Bates College\\
44 Campus Avenue\\
Lewiston, Maine 04210}
\date{}
\maketitle

\begin{abstract}

In this paper, we study Raychaudhuri's equation in the background
of $R + \beta R^2$  gravity with a phenomenological matter ($\rho
\propto a(t)^{-n}$).  We conclude that even though the Strong Energy
Condition (S.E.C.) for Einstein's gravity, which guarantees singularity,
is $n\geq 2$  for $\rho \propto a(t)^{-n}$, a perturbative analysis of
Raychaudhuri's equation in the background of $R + \beta R^2$
gravity reveals that the big bang singularity may not be guaranteed
for $n > 4$.  We derive the following Strong Energy Conditions for $R
+ \beta R^2$ ($\beta \not= 0$):

1) For $k<0$ FRW metric,  S.E.C. is ($0\leq n\leq 4$) \ \  i.e., $-\rho_n \leq
p_n
\leq {1\over 3}\rho_n$.

2) For $k=0$ FRW metric,  S.E.C. is ( $1\leq n\leq 4$)\ \  i.e., $-{2\over
3}\rho_n
\leq  p_n \leq {1\over 3}\rho_n$.

3) For $k>0$ FRW metric,  S.E.C. is ($2\leq n\leq 4$)\ \  i.e., $-{1\over
3}\rho_n
\leq  p_n \leq {1\over 3}\rho_n$.

\end{abstract}

\begin{flushleft}{\large \bf I. INTRODUCTION}
\end{flushleft}
\setcounter{chapter}{1}
\setcounter{equation}{0}

An essential mathematical criterion in the singularity theorem is the
Strong Energy Condition (S.E.C.).  The theorem predicts,
modulo plausible assumptions, in the existence of black holes and
cosmological singularity [1,2]

Singularity theorem is limited by its classical aspect.  The question of
validity of this theorem or how the theorem should be altered when
quantum effects are incorporated is a fascinating question.

A straightforward method of incorporating quantum effects is to
start from the classical Einstein equations and quantize both gravity and
matter via corresponding principle, i.e.,
$\{x,p\}_{poisson} \rightarrow [x,p]_{QM}$.  Specifically, utilizing
Schrodinger representation for canonical momenta results in the
Wheeler-DeWitt equation [3,4].  As for the question of quantum
effects on the big bang singularity, study of the Wheeler-DeWitt
equation has taught us the importance of another parameter, i.e.,
boundary condition (initial condition) for the Wheeler DeWitt
wavefunction[5-8].  The reasoning is quite simple. Quantum mechanical evolution
of a system is inherently a description of a diffusive
system. As a diffusive system, one can only hope to evolve the system
forward in time from some specified initial (big bang) condition on a
Wheeler DeWitt wavefunction. There are also models for which
the resulting Wheeler DeWitt equations become eigenvalue problems
and the arbitrariness of initial conditions is removed [9-11].

Recently, additional method of incorporating quantum effects has
come into focus.  Advances in quantum gravity have revealed that,
even if one starts with the Einstein's gravity coupled to matter, quantum
loop effects of gravity+matter and renormalization procedures result
in a quadratic gravity[12,13].  Therefore, even a classical analysis of
quadratic gravity is inherently a semiclassical analysis.

The two methods of incorporating quantum effects, which differ by
the choice of canonical or covariant quantization, are not necessarily
equivalent.  This is because a formal equivalent of canonical and
covariant quantization is only true for renormalizable theories.

Several authors have studied classical solutions of $R+\beta R^2$ gravity
without matter and  have concluded that Big Bang singularity may be
avoided [14-16].  There are several
problems with higher derivative theories [17,18], e.g., need for additional
initial conditions in a formulation of a cauchy problem, existence of run away
solutions, and whether solutions obtained from $R+\beta R^2$ gravity reduce
to solutions of Einstein's gravity as $\beta \rightarrow 0$.  In order to
resolve these issues, the author has recently proposed an alternate
approach to $R+\beta R^2$ gravity in which $\beta R^2$ is treated as
a backreaction on Einstein's gravity [19].  In essence, in this approach the
gravitational degree of freedom is not altered from those of
Einstein's gravity.

Using such an interpretation of $R + \beta R^2$ gravity, the author also
investigated a classical and Wheeler
DeWitt evolution of $R + \beta R^2$
 gravity for a particular sign of $\beta$, corresponding to non-
tachyon case.  Matter was described by a phenomenological $\rho
\propto a(t)^{-n}$.  It was concluded that both the Friedmann
potential $U(a)$ ($ {\dot a}^2 + 2U(a) = 0 $) and the Wheeler DeWitt
potential $W(a)$ ($\left[-{\partial^2\over \partial a^2} +
2W(a)\right]\psi (a) =0 $)  develop repulsive barriers near
$a\approx 0$  for $n>4$  (i.e., $ p > {1\over 3}\rho $). The
interpretations was clear.  Repulsive barrier in $U(a)$ implies that a
contracting FRW universe ($k>0, k=0, k<0$)  will bounce to an
expansion phase without a total gravitational collapse. Repulsive
barrier in $W(a)$ means that $a \approx 0$  is a classically forbidden
region.  Therefore, probability of finding a universe with the big bang
singularity ($a=0$ ) is exponentially suppressed.

Superficially, the prediction of no cosmological singularity for $n >
4$ (i.e., $ p > {1\over 3}\rho $) seems to be in violation of the Singularity
Theorem for Einstein's gravity, which predicts singularity for a matter
satisfying the S.E.C., i.e., $ p \geq -{1\over 3}\rho $
and $ p \geq -\rho $ or equivalently $n\geq 2$ for $\rho \propto a(t)^{-n}$
[20].

In this paper, we study Raychaudhuri's equation in the background
of $R + \beta R^2$ gravity coupled to matter ($\rho \propto a(t)^{-
n}$).  We conclude that the appropriate S.E.C. for $R + \beta R^2$ gravity is
different from that of Einstein's gravity.  We also derive explicit
expressions for S.E.C.  for $R+ \beta R^2$ gravity.  The S.E.C. for $R + \beta
R^2$ gravity is shown to be
in agreement with the author's previous work, which demonstrated
that both classical and Wheeler DeWitt solutions  of $R + \beta R^2$ gravity
were free of cosmological singularity for $n> 4$.

Sign conventions used in this paper are as follows. $g=(-,+,+,+)$, \ \
$R_{ab} -{1\over 2}g_{ab}R = \left (+ \right )8\pi
GT_{ab}$, \ \ $\left (+ \right ) R(U,V) =
\nabla_{U}\nabla_{V}- \nabla_{V}\nabla_{U}-
\nabla_{\left[U,V\right]}$.

\begin{flushleft}{\large \bf II. REVIEW OF STRONG ENERGY
CONDITION FOR EINSTEIN'S GRAVITY
}
\end{flushleft}
\setcounter{chapter}{2}
\setcounter{equation}{0}

Even though the goal of this paper is to study Strong Energy
Condition (S.E.C.) for a quadratic gravity, much of the technique
will be borrowed from the analysis of S.E.C. for Einstein's gravity.
Therefore, we shall briefly review the derivation of S.E.C. for
Einstein's gravity. For a pedagogical review, please see [20].

	Let $\xi^a$ be a tangent vector for a congruence of
timelike geodesics.  For a hypersurface orthogonal congruence,
Raychaudhuri's equation is

\begin{equation}
{d\theta\over d\tau} = -{1\over
3}\theta^2 - \sigma_{ab}\sigma^{ab} + R_{ab}\xi^a\xi^b.\end{equation}

\noindent
$\theta$  and $\sigma^{ab}$, are the expansion and sheer of
two nearby tangent vectors, respectively.  In Einstein's gravity,
S.E.C. and the ensuing singularity theorem, follow from
requiring that
\begin{equation} -R_{ab}\xi^a\xi^b = 8\pi G\left[ T_{ab} -
{1\over 2} Tg_{ab}\right]\xi^a\xi^b  \geq 0\end{equation}

\noindent
for all timelike $\xi^a$.
In such a case, ${d\theta\over d\tau} <0$  and a pair of nearby
timelike geodesic vectors converges and will eventually intersect.

We will be interested in the cosmological singularity.  In a FRW
metric, matter is described by

\begin{equation}
T^a_b = \left( \begin{array}{cccc}
-\rho &&& \\
& p &&\\
&& p & \\
&&& p \end{array} \right) = \rho t^at_b + p x^ax_b + p y^ay_b+
p x^az_b.
\end{equation}

\noindent
$\{t^a, x^a, y^a, z^a\}$ are eigenvectors of $T^a_b$. They are
normalized as $-t^at_a = x^ax_a=y^ay_a=z^az_a =1$.

\noindent Because of isotropy, one can always rotate the coordinate such that
the most general forward timelike vector is

\begin{equation}
\xi^a = At^a + \left( A^2 -1\right)^{1/2}x^a, \ \ (A\in
(1,\infty ), \ \ \ \xi^a\xi_a = -1).
\end{equation}

\noindent Then,

\begin{equation}
-R_{ab}\xi^a\xi^b = 8\pi G\left[ T_{ab} - {1\over 2}
Tg_{ab}\right]\xi^a\xi^b =
8\pi G \left [ \rho \left( A^2 - {1\over 2}\right)  +p\left( A^2 +
{1\over 2}\right) \right].
\end{equation}

\noindent S.E.C. is a requirement on an equation of state $(p,\rho)$ for which
$-R_{ab}\xi^a\xi^b \geq 0$  is true for all timelike vectors $\xi^a$, i.e.,
$A\in (1,\infty )$.

Expression (2.5) is monotonic function of $A^2$.  Therefore, if we
can find an equation of state  $(p,\rho)$ for which $-
R_{ab}\xi^a\xi^b \geq 0$ is satisfied by two extreme timelike
vectors ($A=1$ and $A \rightarrow \infty$), then this $(p,\rho)$ is
guaranteed to satisfy $-R_{ab}\xi^a\xi^b \geq 0$ for all timelike
vectors  $A\in (1,\infty )$.  Requiring $-R_{ab}\xi^a\xi^b \geq 0$ for $A=1$
gives $
\rho + 3p \geq 0$.  Requiring $-R_{ab}\xi^a\xi^b \geq
0$ for $A \rightarrow \infty$  gives $ \rho + p \geq 0$.  These
are the familiar S.E.C. [20].

\begin{flushleft}{\large \bf III. DERIVATION OF STRONG ENERGY
CONDITION FOR $R + R^2$ GRAVITY}
\end{flushleft}
\setcounter{chapter}{3}
\setcounter{equation}{0}

Raychaudhuri's equation describes how a congruence of
timelike geodesics deviate from one another.  Therefore,
Raychaudhuri's equation  (2.1) is valid even in a background of
quadratic gravity.   As in Einstein's gravity, one can be assured of
a convergence of congruence of timelike geodesics by requiring
$-R_{ab}\xi^a\xi^b \geq 0$.  In order to proceed, we need the relevant
``Einstein" equations for
quadratic gravity.  A pedagogical derivation is given in Appendix.

\begin{equation} {1\over2}Rg_{ab} - R_{ab}+16\pi
G\beta\left({1\over 2}R^2g_{ab} - 2RR_{ab} +
2R_{;n}^{;n}g_{ab} - 2R_{;a;b}\right) = 8\pi
GT_{ab}.\end{equation}

\noindent The trace of this equation is
\begin{equation} 6\cdot 16\pi G\beta R_{;n}^{;n} + R =
8\pi GT.\end{equation}

Some comments are in order.  First, by dimensional
consideration, $\beta$ is dimensionless.  Second, as noted by
various authors [21,22], (3.2) resembles a scalar field equation with
$m^2 = -(6\cdot 16 \pi G\beta)^{-1}$.   Therefore, $\beta <0$
is needed to eliminated tachyons.  Third, an order of magnitude
estimate reveals that contributions from the  quadratic terms
are smaller than those of Einstein's terms by $\beta GR \approx \beta G^2\rho
\approx \beta \rho /
\rho_{planck} $ .  Therefore, if we make a reasonable
assumption that $\beta \approx  O(1)$, than we will be justified
in treating the quadratic terms perturbatively until the very
early universe. For a more elaborate discussion on the justification
for treating the $\beta R^2$ as a perturbation, please see [19].
The subsequent analysis should be
viewed as a perturbative probe into a possible nonlinear
phenomenon.
On the other hand, it is worth noting that Mijic[23], Starobinsky[24],  and
Berkin[25] have
studied the large $\beta$ range and concluded that even for a
vacuum, gravity alone can generate an inflationary phase in $R +
\beta R^2$ gravity.

For use in Raychaudhuri's equation, we need an expression for
Ricci tensor $R_{ab}$    in terms of $T_{ab}.$

\noindent Combining (3.1) and (3.2) we get
\begin{equation}
- R_{ab}+16\pi
G\beta\left({1\over 2}R^2g_{ab} - 2RR_{ab} -R_{;n}^{;n}g_{ab} -R_{;a;b}\right)
= 8\pi
G(T_{ab} - {1\over 2}Tg_{ab}).
\end{equation}

\noindent  We are interested in the first order contribution from $\beta R^2$
to
Raychaudhuri's equation.  Therefore, for terms already
multiplied by $\beta$ in (3.1), we may substituting $R_{ab} = -8\pi
G\left( T_{ab} -{1\over 2}Tg_{ab}\right) + O(\beta) $ and
$R=8\pi GT + O(\beta)$ to get

\begin{equation}
\left({1\over 2}R^2g_{ab} - 2RR_{ab} -
2R_{;n}^{;n}g_{ab} - 2R_{;a;b}\right) =
-\tilde G\left({1\over 2}T^2g_{ab}\tilde G -2TT_{ab}\tilde G +
T_{;n}^{;n}g_{ab} + 2T_{;a;b}\right) + O(\beta).
\end{equation}

\noindent We have introduced a notation $\tilde G \equiv  8\pi G$.
And finally, for use in Raychaudhuri's equation, we get

\begin{equation} -R_{ab}\xi^a\xi^b = \left[
\begin{array}{ll}\tilde G(T_{ab} - {1\over 2}Tg_{ab}) + &\\
2\beta{\tilde G}^2 \left ( {1\over 2}T^2g_{ab}\tilde G -2TT_{ab}\tilde
G + T_{;n}^{;n}g_{ab} + 2T_{;a;b}\right)  & \end{array} \right]
\xi^a\xi^b+ O(\beta^2).
\end{equation}

\noindent The rest of the procedure more of less mimicks  Einstein's
case.   Expression for a general timelike vector $\xi^a$ is
(2.4).   $T_{ab}$ appropriate for a FRW metric is (2.2).

We will assume that during any epoch in the evolution of
universe, the universe is dominated by a matter with a
characteristic dependence on the scale factor (i.e., $\rho =
{\rho_0\over a^n} $).   A local  conservation of $T_{ab}$ gives $p
= {n-3\over 3}{\rho_0\over a^n}$  .

A purist may argue that a proper way of including matter in the
early universe is by incorporating a quantum field (e.g., scalar field).
We feel that in the study of Raychaudhuri's equation, describing
matter by a phenomenological $\rho = {\rho_0\over a^n} $
  is adequate.  This is because Raychaudhuri's equation studies
geometrical optics limit of a field in a background metric and
matter.  If we had used a scalar field, than we would have had
a difficult task of splitting a field $\phi (x,t) = \phi_0 (x,t)+
\delta\phi (x,t)$, where $\phi_0 (x,t)$ and $\delta\phi (x,t)$
are a background low frequency and a fluctuating high frequency
components, respectively.  The background
component $\phi_0 (x,t)$ would again result in an effective
$\rho = \langle T^0_0 [\phi_0 (x,t)]\rangle_{space} \propto a^{-
n}$  for some ($n$). The value of ($n$) would  depend on properties of the
field (e.g.,
mass, self coupling $V(\phi )$). And Raychaudhuri's equation
would correspond to a geometrical optics limit of $\delta\phi
(x,t)$.

Continuing, it is a laborious exercise to show that (3.5) becomes

\begin{equation}
-R_{ab}\xi^a\xi^b= \begin{array}{cc}
{n-2\over 2}\tilde G\rho_n + \beta{\tilde G}^2\left[ 3n(n-1)(n-4)
\tilde G \rho_n^2  - 6 n^2(n-4)ka^{-2}\rho_n\right] + &  {} \\
(A^2 -1) \left[ {n\over 3}\tilde G\rho_n + \beta{\tilde G}^2\left[
2n^2(n-4) \tilde G \rho_n^2  - 4 n(n+2) (n-4)ka^{-2}\rho_n\right]
\right] & {}
\end{array}
\end{equation}

A reader who may want to derive (3.6) from (3.5) will find
the following helpful.
\begin{equation} T \equiv T_{ab}g^{ab} = 3p-\rho = (n-
4)\rho_n\end{equation}

\begin{equation} T_{ab}\xi^a\xi^b = A^2(p+ \rho)
-p\end{equation}

\begin{equation} T_{;a;b}\xi^a\xi^b = A^2\left( \partial_t^2T
- {\dot a\over a}\partial_tT\right) + {\dot a\over
a}\partial_tT\end{equation}

\begin{equation}
T^{;a}_{;a} = - a^{-3}\partial_t\left( a^3\partial_tT\right) =- (n-
4) a^{-3}\partial_t\left( a^3\partial_t\rho_n\right)
\end{equation}

Again, Strong Energy Condition (S.E.C.) is requirement on
equation of state for which  $-R_{ab}\xi^a\xi^b \geq 0$ for
all forward timelike vector $\xi^a$.  There is
a subtlety to keep in mind.  From (2.1), even though $-
R_{ab}\xi^a\xi^b \geq 0$ implies that a pair of timelike
geodesics converge, the opposite  $-R_{ab}\xi^a\xi^b \leq 0$ does
not imply that a pair of timelike geodesics diverge.  This is
because of existence of other negative negative terms in (2.1)

Expression (3.6) is a monotonic function of $A^2$.  Similar to
Einstein's case, if we can find  an equation of state  $(p,\rho)$  for
which $-R_{ab}\xi^a\xi^b \geq 0$ is satisfied by two
extreme timelike vectors ($A=1$ and $A \rightarrow \infty$
), then this $(p, \rho)$  is guaranteed to satisfy $-
R_{ab}\xi^a\xi^b \geq 0$ for all timelike vectors  $A\in
(1,\infty )$.

\noindent Requiring $-R_{ab}\xi^a\xi^b \geq 0$ for $A=1$ gives

\begin{equation} {n-2\over 2}\tilde G\rho_n + \beta{\tilde
G}^2\left[ 3n(n-1)(n-4) \tilde G \rho_n^2  - 6 n^2(n-4)ka^{-
2}\rho_n\right]  \geq 0. \end{equation}

\noindent Requiring $-R_{ab}\xi^a\xi^b \geq 0.$ for $A \rightarrow \infty$
gives

\begin{equation}
{n\over 3}\tilde G\rho_n + \beta{\tilde G}^2\left[ 2n^2(n-4) \tilde
G \rho_n^2  - 4 n(n+2) (n-4)ka^{-2}\rho_n\right] \geq
0.\end{equation}

\noindent The problem of cosmological  S.E.C. has been reduced to solving
for ($n$) that  will satisfy (3.11) and (3.12) as $a \rightarrow 0$.

As a partial check in algebra, consider the case of $\beta =0$.
(3.11) gives $n \geq 2$ , which is equivalent to $p_n > -{1\over
3}\rho_n$.  And (3.12) gives $n \geq 0$, which is equivalent to $p_n >
-\rho_n$.  As expected, these are the S.E.C. for Einstein's gravity.

Now for $\beta \not= 0$  case ($\beta <0$  for a tachyon free
system), the argument is significantly more subtle.
We note that as one goes further back in time, depending on the
value of
($n$), one of the ($\rho_n, \rho^2_n, a^{-2}\rho_n$)  in (3.11)
and (3.12) will grow fastest and hence dominate the expressions.

We already know the result if $\rho_n$ dominates (i.e., $\beta=0$).
We need $n \geq 2$ to satisfy (3.11) and (3.12) (Figure 1a).

Now if $\rho^2_n$ dominates, than (3.12) reduces to  $\beta n (n-
1)(n-4) \geq 0$,  i.e.,  $1\leq n \leq 4$  since $\beta <0$ .   And
(3.12)  reduces to $\beta n^2(n-4) \geq 0$,  i.e., $0\leq n \leq 4$.
The two are simultaneously satisfied by $1\leq n \leq 4$ (Figure 1b).

And finally, if the curvature term ($ ka^{-2}\rho_n$) dominates,
than (3.11)  reduces to  $-k\beta n^2(n-4) \geq 0$ and (3.12)
reduces to $-k\beta n (n+2)(n-4) \geq 0$.  The two reduced
equations are satisfied simultaneously for
($k>0$) FRW metric by $n \geq 4$ (Figure 1c), and  for ($k<0$) FRW metric by
$0\leq n \leq
4$   (Figure 1d).  We have tactfully assumed that $n \geq 0$, i.e.,
energy density should not get less dense when squeezed.

 We also note  that only when $0\leq n
\leq 2$  can the curvature term $ka^{-2}\rho_n$ grow to  dominate
over the $\rho^2_n$ term, as one goes further back in time ($a \rightarrow 0$)

We have obtained various collection of results.  A slightly more
useful conclusion would be, for a given ($\rho_n, k, \beta$)
whether there will be a big bang singularity.  We will proceed
to address this question in two steps.  The strategy will be as follows.
First, for a given ($\rho_n, k, \beta$) we will have to determine
which of the ($\rho_n, \rho^2_n, a^{-2}\rho_n$) will dominate (3.11, 3.12))  as
$a\rightarrow 0$ .  Then from Figures 1b-1d, we will be able to read off
whether such a given set of ($\rho_n, k, \beta$) has a big bang singularity.

Let us first consider $k=0$ FRW metric, with $\rho \propto a(t)^{-
n}$.   As one goes back to earlier time, $\rho^2_n$ should
dominate over $\rho_n$.  Therefore, from (Figure 2a) the appropriate S.E.C. is
($1\leq n\leq 4$).

A careful combination of previous results reveals that the
 S.E.C. for $k<0$ FRW metric is $0\leq n \leq 4$  (Figure 2b).
The reasoning is as follows. As $a\rightarrow 0$, either  $ka^{-
2}\rho_n$ or $\rho^2_n$ could end up dominating (3.11, 3.12).  For  $0\leq n
\leq 2$, $ka^{-2}\rho_n$ will dominate.  And
from Figure 1d, all of $0\leq n \leq 2$  results  in a convergence of
a pair of timelike vectors.  For $n \geq 2$, $\rho^2_n$ will
end up dominating.  From Figure 1b, of this region, only
$2\leq n \leq 4$  results in a convergence of a pair of timelike
vectors Q.E.D.

Similarly, analysis reveals that the S.E.C. for $k>0$ FRW
universe is $2\leq n \leq 4$ (Figure 2b).  Again for  $0\leq n
\leq 2$ , $ka^{-2}\rho_n$ will dominate. But from Figure 1d,
none  of this region results in a convergence of a pair of timelike vectors.
For $n \geq 2$, $\rho^2_n$ will again end up
dominating.  And from Figure 1b, of this region, only $2\leq n
\leq 4$  results in a convergence of a pair of timelike vectors
Q.E.D.

These results, of cosmological strong energy condition for
$R+R^2$ gravity, are summarized in Figures 2a-2c.

\begin{flushleft}{\large \bf IV. DISCUSSIONS AND CONCLUSION}
\end{flushleft}
\setcounter{chapter}{4}
\setcounter{equation}{0}

 We were primarily interested in how Big Bang singularity
would be affected by  quantum effects.  Granted
that even a classical analysis of  $R+\beta R^2$ gravity can be interpreted as
a
semiclassical analysis, there are several limitations to our
analysis and we would like to briefly raise these points.

1) Near the Planck epoch, large quantum fluctuations will undoubtedly
result in an inhomogeneous universe, yet we have assumed
homogeneous (and isotropic) metric in (2.3-2.4) to make the
problem analytically tractable. A future work will have to
address how the present conclusion is affected by
anisotropy and inhomogeneity.

2) Raychaudhuri's equation is a classical geometric optics limit,
which may be invalid near the Planck epoch.

3) In Raychaudhuri's equation, we have taken the limit as
$a\rightarrow 0$.  A more physical limit might be to cut off
the limits at $\rho_n \rightarrow \rho_{planck}, \ \ ka^{-2}
\rightarrow (planck \ \ length)^{-2}$.  Various set of conclusions obtained by
this limiting process
turns out to be sensitive to fine tuning of parameters and will
not be discussed here. An interested reader is invited to
explore various possibilities.

4) In retrospect, in the deriving of S.E.C. for $R+\beta R^2$ gravity, we have
pushed the analysis until $k\beta a^{-2}\rho_n$ or
$\beta\rho^2_n$  dominated over $\rho_n$.  Yet, if the terms
linear in $\beta$ ever grow larger than the terms from Einstein's
gravity, than this will be outside the validity of perturbation.
This is a serious objection.  But as it is a common practice in perturbation
theory, we are inclined to interpret these results obtained by a
perturbative analysis as a possible glimpse into
a nonperturbative effect.  With this in mind, we must settle for a
following weaker conclusion.

``Cosmological S.E.C. for Einstein gravity is $n\geq 2$ .  For
$R + \beta R^2$ gravity,  first order perturbation analysis
indicates a deformation of S.E.C. region such that:

1) for $k<0$, $n\geq 4$ may have to be  excluded and   $0\leq n
\leq 2$  may have to be included.

2) for $k=0$, $n\geq 4$ may have to be  excluded and   $1\leq n
\leq 2$  may have to be included.

3) for $k>0$, $n\geq 4$  may have to be excluded."

In closing, these results compliment the authors recent work, which
demonstrated that both classical and Wheeler DeWitt solutions
of $R + \beta R^2$ gravity were free of big bang singular free for $n >
4$.

\begin{flushleft}{\large \bf V. APPENDIX}
\end{flushleft}
\setcounter{chapter}{5}
\setcounter{equation}{0}

The action for quadratic gravity is  \begin{equation} I = -{1\over
16\pi G}\int d^4x\sqrt{-g}R -\int d^4x
\left
[\beta_1R^2+\beta_2R_{ab}R^{ab}+\beta_3R_{abcd}R^{abcd}\right] +
I_{matter} + {\rm surface \; term}.\end{equation} We have formally included a
surface term to cancel any boundary
term that would result in applying the variational principle.
We will be interested in applying the formalism to homogeneous and
isotropic metric, i.e., Wely tensor vanishes $C_{abcd}=0$\ [26].
By definition of Weyl tensor, $C_{abcd}C^{abcd}=R_{abcd}R^{abcd}
-2R_{ab}R^{ab} +{1\over 3}R^2$.
This gives one relationship among the possible quadratic terms.

\noindent The second relationship is from the 4 dimensional generalization of
Gauss Bonnet formula [12],

\begin{equation} R^2 - 4R_{ab}R^{ab} + R_{abcd}R^{abcd} = {\rm exact
\; derivative.}\end{equation}

\noindent The two relationships, combined with the fact that Euler Lagrange
equations are unchanged by addition of an
exact differential, allow any two of
$\beta_1, \beta_2, \beta_3$ to be set equal to zero in the action
(5.1).  We choose to set $\beta_3=\beta_2=0$.

The resulting Euler Lagrange equations are
\begin{equation} {1\over2}Rg_{ab} - R_{ab}+16\pi
G\beta\left({1\over 2}R^2g_{ab} - 2RR_{ab} +
2R_{;\sigma}^{;\sigma}g_{ab} - 2R_{;a;b}\right) = 8\pi
GT_{ab}.\end{equation}

\noindent The trace of this equation is
\begin{equation} 6\cdot 16\pi G\beta R_{;\sigma}^{;\sigma} + R =
8\pi GT.\end{equation}


\newpage

\begin{center}
{\large \bf FIGURE CAPTIONS}
\end{center}
\begin{enumerate}

\item Figure a: Strong Energy Condition for Einstein's gravity (i.e., $\beta
=0$).
\newline
Figure b:  $-R_{ab}\xi^a\xi^b \geq 0$ determined under the
assumption of $\rho_n^2$ domination.
\newline
Figure c:  $-R_{ab}\xi^a\xi^b \geq 0$ determined under the
assumption of $ka^{-2}\rho_n$ domination ($k>0$).
\newline
Figure d:  $-R_{ab}\xi^a\xi^b \geq 0$ determined under the
assumption of $ka^{-2}\rho_n$ domination ($k<0$).
\newline
\item Figure a: Strong Energy Condition for $k=0$ FRW metric, ($1\leq
n\leq 4$ i.e., $-{2\over 3}\rho_n \leq  p_n \leq {1\over 3}\rho_n$ ). For
$1\leq n\leq 4$, a universe ends in $\rho_n^2$ domination.
\newline
Figure b: Strong Energy Condition for $k<0$ FRW metric, ($0\leq
n\leq 4$ i.e., $-\rho_n \leq  p_n \leq {1\over 3}\rho_n$).   For $0\leq n\leq
2$, a universe ends in $ka^{-2}\rho_n$ domination.  For $2\leq n\leq 4$, a
universe ends in $\rho_n^2$ domination.
\newline
Figure c: Strong Energy Condition for $k>0$ FRW metric, ($2\leq
n\leq 4$ i.e., $-{1\over 3}\rho_n \leq  p_n \leq {1\over 3}\rho_n$). For $2\leq
n\leq 4$, a universe ends in $\rho_n^2$ domination.

\end{enumerate}

\newpage

\begin{center}
{\large \bf REFERENCES}
\end{center}
\begin{enumerate}

\item S.W. Hawking, Proc. Roy. Soc. Lon. {\bf A300}, 187, (1967).

\item S.W. Hawking, Proc. Roy. Soc. Lon. {\bf A314}, 529, (1970).

\item B. S. DeWitt, Phys. Rev. {\bf 160}, 1113 (1967).

\item J. A. Wheeler, in {\it Battelle Rencontres}, edited by C. DeWitt
and
J. A. Wheeler (Benjamin, New York, 1968).

\item J. B. Hartle and S. W. Hawking, Phys. Rev. {\bf D 28}, 2960
(1983).

\item S. W. Hawking, Nucl. Phys. {\bf B 239}, 257 (1984).

\item A. Vilenkin, Phys. Rev. {\bf D 33}, 3560 (1986).

\item A. Vilenkin, Phys. Rev. {\bf D 37}, 888 (1988).

\item J.H. Kung, Gen. Rel. Grav. {\bf 27}, 35 (1995).

\item T. Padmanabhan, Gen. Rel. Grav. {\bf 14}, 549 (1982).

\item M. Cavaglia, V. de Alfaro, and A. Filippov, Int. Jou. Mod. Phys.
{\bf A10}, n.5 (1995).

\item S. Deser, in {\it Quantum Gravity}, edited by C. J. Isham, R.
Penrose and D. W. Sciama, (Oxford University Press, 1975).

\item R. Utiyama and B.S. DeWitt, J. Math. Phys. {\bf 3}, 608 (1962).

\item V. Muller and H. J. Schmidt, Gen. Rel. Grav. {\bf 17}, 769
(1985).

\item D. Page, Phys. Rev. {\bf D 36}, 1607 (1987).

\item J.D. Barrow and A. Ottewill, J. Phys. A: Math. Gen. {\bf 16},
2757 (1983).

\item X. Jaen et al., Phys. Rev. {\bf D 34}, 2302 (1986).

\item A. Strominger, in {\it Quantum Theory of Gravity}, edited by S.
Christensen (Adam Hilger, 1984).

\item J. H. Kung, ``$R+ R^2$ Gravity as $R+$ Backreaction", to appear in Phys.
Rev. D. (gr-qc 9509058).

\item S.W. Hawking and G. F. R. Ellis, {\it The Large Scale Structure of
Space-Time} (Cambridge University Press, 1973).

\item B. Whitt, Phys. Lett. {\bf 145B}, 176 (1984).

\item J. C. Alonso, F. Barbero, J. Julve, and A. Tiemblo, Class. Quantum
Grav. {\bf 11}, 865 (1994).

\item M. Mijic et al., Phys. Rev. {\bf D 34}, 2934 (1986).

\item A. A. Starobinsky and H. J. Schmidt, Class. Quantum Grav. {\bf
4}, 695 (1987).

\item A. L. Berkin, Phys. Rev. {\bf D 42}, 1016 (1990).

\item R.M. Wald, {\it General Relativity} (The university of Chicago Press,
1984).

\end{enumerate}

\end{document}